\documentstyle[prl,aps,twocolumn,epsf]{revtex}
\def\break#1{\pagebreak \vspace*{#1}}
\def\epsfig#1#2#3#4
         {
         \epsfysize=#2 \vbox{ \hglue#3 \epsfbox[#4]{#1} }
         }
\def\epsfigrotl#1#2#3#4
         {
         \epsfxsize=#2 
         \setbox\rotbox=\hbox to #2{\epsfbox[#4]{#1}}
         \vbox{\hglue#3 \rotl\rotbox}
         }
\def\epsfigrotr#1#2#3#4
         {
         \epsfxsize=#2 
         \setbox\rotbox=\hbox to #2{\epsfbox[#4]{#1}}
         \vbox{\hglue#3 \rotr\rotbox}
         }
\newbox\rotbox
\input rotate

\newcommand{\B}{B}
\newcommand{\scsc}{\rm}

\begin{document}

\title{A Dynamical Theory of Electron Transfer: 
Crossover from Weak to Strong Electronic Coupling}

\author{J\"urgen T. Stockburger and C.H. Mak}

\address{Department of Chemistry, University of
Southern California, Los Angeles, CA 90089-0482}

\date{Date: July 12, 1996}

\maketitle

\widetext
\begin{abstract}
We present a real-time path integral theory for the rate of electron
transfer reactions.  Using graph theoretic techniques, the dynamics 
is expressed in a formally exact way as a set of integral equations.
With a simple approximation for the self-energy, the rate can then be 
computed analytically to all orders in the electronic coupling matrix 
element.  We present results for the crossover region between weak 
(nonadiabatic) and strong (adiabatic) electronic coupling and show that 
this theory provides a rigorous justification for the salient features 
of the rate expected within conventional electron transfer theory.  
Nonetheless, we find distinct characteristics of quantum behavior 
even in the strongly adiabatic limit where classical rate theory is 
conventionally thought to be applicable.
To our knowledge, this theory is the first systematic dynamical treatment 
of the full crossover region.

\end{abstract}

\narrowtext

\section{Introduction}

Since the landmark papers of Marcus \cite{Marcus,reviewET} on the theory of 
electron transfer (ET) reactions, the central role of the solvent in
determining the ET rate has been well recognized.
The ordinary ET reaction is an activated process \cite{reviewET}, 
in which there is a free energy barrier separating reactants and
products.  This barrier is the result of the often strongly solvated 
donor and acceptor states, and the transfer of an electron then 
requires a specific large-scale reorganization of the 
environment, usually achieved through equilibrium fluctuations.

ET reactions are commonly thought to be well understood 
in two distinct limits (see Fig.~\ref{intro}).  First, for a very weak 
electronic coupling or the so-called nonadiabatic limit, the ET 
can be treated with a perturbation theory \cite{reviewET}, 
which to lowest order in the electronic coupling
leads to the Golden Rule expression for the rate.
For the opposite adiabatic limit in
which the electronic coupling is large, the rate is thought
to be controlled by the motion of the solvent on the lower 
electronic state and well described by classical activated
rate theory \cite{reviewET}.
In both nonadiabatic and adiabatic theories, there is an assumed 
separation of timescales between the motions of the electron and
the solvent.  
In the nonadiabatic limit, the solvent fluctuations
are assumed fast and the ET is controlled by
thermal excitations of the solvent to near the crossing region followed
by a fast transversal of the crossing region by the solvent.
The ET rate is therefore controlled by the probability of 
an electronic transition 
proportional to the square of the electronic coupling.
The adiabatic limit on the other hand is characterized by 
slow solvent fluctuations such that the Born-Oppenheimer approximation
for the electron can be invoked.
Therefore according to conventional theories in these two limits, 
the ET rate is expected to depend on the electronic coupling $\hbar\Delta/2$ 
like
\break{1.5in}
\begin{equation}
k_{\rm ET} \sim \left\{ \begin{array}{ll}
                          \Delta^2 \exp (-E_0/kT),   & \mbox{nonadiabatic} \\
                           \exp (-[E_0-\hbar\Delta/2]/kT), & \mbox{adiabatic} 
                         \end{array}
                 \right.
\end{equation}
where $T$ is the temperature and $k$ is the Boltzmann constant.

Regarding the assumed separation of timescales, there are at 
least two aspects of ET reactions which remain largely unclear and demand 
more complete theoretical treatments.
First, there is the question of the ``crossover region''.
By this we mean the region between the nonadiabatic and the adiabatic
limits in which the electronic matrix element changes from the 
weak coupling to the strong coupling limit.  In this region, the
assumed separation of timescales is necessarily violated.
In fact, a successful dynamical crossover theory that can unify the two 
limits requires inclusion of high order electronic coupling terms
\cite{garg,jortner,mukamel},
and such a theory has yet to be found.  
Second, as is true in almost all condensed phase ET reactions, the frequency
spectrum of solvent motions is charac-
\vbox{
\epsfverbosetrue
\begin{figure}
\epsfxsize=\columnwidth \epsfbox[85 70 545 300]{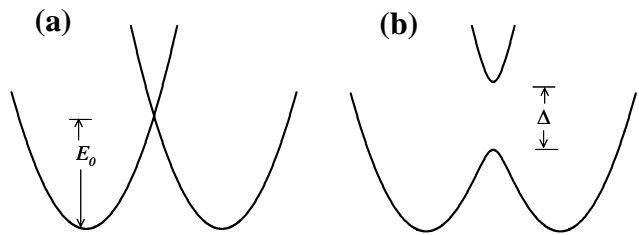}
\caption[]
{\label{intro}
Potential surfaces for nonadiabatic (a) and adiabatic (b)
electron transfer. In the nonadiabatic case, the Landau-Zener
mechanism facilitates transfer at the crossing point of two diabatic
surfaces. The adiabatic transfer is often described as a barrier
crossing on the lower adiabatic energy surface.
}
\end{figure}
}
terized by a broad distribution
of timescales.  Hence, the assumption of timescale separation, while 
perhaps true for parts of the spectrum, may not be so for others.
While it is commonly believed that fast solvent
polarization has little effect on the ET rate \cite{reviewET,marcus65}, 
unconventional behaviors have been suggested for solvents with
high-frequency modes \cite{kim-hynes}.
Indeed, Stuchebrukhov and Song \cite{stuchebrukhov} have recently reported 
that high-frequency components in the solvent spectrum can substantially 
renormalize the intrinsic electronic 
coupling to slow down the otherwise adiabatic rate.  The same adiabatic 
renormalization effects have also been recognized by others in a somewhat 
different context \cite{bray,leggett}.  The questions surrounding
the influence of high-frequency modes in adiabatic reactions seem far
from settled.

The search for a unified treatment that can connect the conventional
nonadiabatic and adiabatic theories presents a considerable challenge.  
Early attempts were made by Levich \cite{Levich} and Zusman \cite{Zusman}
to use the Landau-Zener formula to incorporate higher-order effects 
in the electronic coupling into the ET rate.  Along a similar line of
thinking, Garg, Onuchic and Ambegaokar \cite{garg} used an approximate 
real-time path integral treatment in an attempt to account for adiabaticity
effects.  

More recent approaches \cite{gehlen1,gehlen2,stuchebrukhov,cao} have
been largely based on an imaginary-time formulation of the problem.
Most noticeable of these is the centroid free energy method
\cite{centroid} employed by Gehlen et al. \cite{gehlen1,gehlen2}.
Extensions of this to complex-valued centroid coordinates by
Stuchebrukhov and Song \cite{stuchebrukhov} have also been useful for
obtaining results for the inverted region where nuclear tunneling is
crucial.  Another imaginary-time theory which does not explicitly
employ the centroid formulation has been given by Cao et al.
\cite{cao} who focus on the computational aspects of instanton
methods.  All these theories rely on an assumed relationship between
the ET rate and an analytic continuation of the partition function for
the tunneling system \cite{Langer}. This relationship, however has
not been derived from first principles for the ET problem.

In this paper, we take a rather different approach to the problem.
With a spin-boson model for the ET reaction, our method sets out from
an exact path integral formulation \cite{leggett,weiss} for the
real-time {\em dynamics} of the electron.  From the path integrals, it
is then easy to see that the ET rate is related to clusters of
kink/anti-kink pairs (called ``blips'') \cite{leggett} which interact
with each other through a nonlocal influence functional.  When the
fluctuations of the bath polarization are sufficiently fast compared
to the motions of the electron, these interblip interactions vanish.
In this case, the rate reduces to nonadiabatic theory and becomes
completely equivalent to a low-order perturbation theory in the
electronic coupling.  On the other hand, when the bath motions are
sufficiently slow compared to the electron, interblip interactions
force blips to bunch up into clusters.  In each cluster, multiple
electronic transitions are possible, and the rate is now given by a
series containing high-order terms in the electronic coupling.  Using
standard graph theoretic techniques and a simple approximation for the
self-energy (i.e., the sum over all irreducible diagrams), we are able
to sum up an infinite number of terms in this series and obtain an
expression for the ET rate not only in the nonadiabatic and adiabatic
limits, but for the entire crossover region.  To our knowledge, this
is the first successful real-time formulation that can span the whole
crossover region.

Section \ref{spinboson} gives a brief overview of the path-integral
formulation of the dynamics of the spin-boson system. With minor
modification, this material follow closely reviews by Leggett
\cite{leggett} and Weiss \cite{weiss}. In Section \ref{dyntransfer},
the formal expression for the time-dependent dynamics is transformed
diagrammatically to yield a rigorous result for the ET rate.  Sections
\ref{nonadrate} and \ref{exactrate} give a pictorial interpretation of
the results, and Section \ref{adcorr} discusses general ET behaviors in
the two limits in terms of topological features of the graphs.
Section \ref{crossover} presents a simple approximation which allows
us to resum the entire perturbation series for the entire crossover
region.  Some numerical results for the crossover behavior are also
discussed.  Finally, section \ref{summary} concludes this paper with a
brief summary of the most important features of the present theory and
our results.

\section{Dynamics of Electron Transfer in the Spin-Boson Model}
\label{spinboson}
For the discussion of ET dynamics in a donor-acceptor system
in the condensed phase, we take the simple spin-boson model
which has been used in many previous studies.  The Hamiltonian is
\begin{equation}
H = H_{\scsc S} + H_{\scsc I} + H_{\scsc B}\;,
\end{equation}
where 
\begin{eqnarray}
H_{\scsc S} &=& -{\hbar\over 2}\left(\Delta\sigma_x + \epsilon\sigma_z\right) \;, \\
H_{\scsc I} &=& -{a\over 2}\sigma_z \sum_\alpha c_\alpha x_\alpha
 \label{HISB} \;, \\
H_{\scsc B} &=& {1\over 2}\sum_\alpha {p_\alpha^2\over m_{\alpha}} +
m_{\alpha} \omega_\alpha^2 x_\alpha^2 \label{HBSB} \;,
\end{eqnarray}
defining two localized electronic states 
(with a possible asymmetry $\epsilon$) and their overlap in the 
tight-binding approximation with a bilinear coupling to a number of
solvent modes, which are assumed harmonic.

Only two quantities of the oscillator bath are relevant to the dynamics
of the electron.  They are 
the temperature associated with the initial thermal
density matrix of the bath, and its spectral density
\begin{equation}
J(\omega) = {\pi\over 2} \sum_\alpha {c_\alpha^2\over m_\alpha\omega_\alpha}
\delta(\omega-\omega_\alpha)\;.
\end{equation}
A commonly used form of $J(\omega)$, which in the classical limit
describes a frequency-independent friction,
is the {\em ohmic\/} spectral density
\begin{equation}
{a^2\over 2\pi\hbar}\,J(\omega) = \alpha\omega e^{-{\omega / \omega_c}}\;,
\end{equation}
where $\alpha$ is a dimensionless parameter which characterizes the
strength of the dissipation.  We shall use
this spectral density throughout this paper.
Classically, the ohmic bath is characterized by a single parameter --
the reorganization energy $\hbar\Lambda = 2\alpha\hbar\omega_c$.

Using standard path-integral techniques
\cite{Feynman-Vernon,leggett,weiss,garg}, a partial trace 
can be performed over the
oscillator coordinates to eliminate them from the problem.  One is left
with an effective 
double path integral expression for the dynamics of the reduced density
matrix for the electron.

The forward- and reverse-time paths for the electron $\sigma(t)$ and
$\sigma'(t)$ are piecewise constant functions of time, alternating
between the two allowed values $\pm 1$.
This particular class of functions can obviously 
be parametrized by the number of
transitions between the two allowed values, 
the times at which these transitions occur, and the constant
values the function assumes in the intervening intervals.

Applying this parametrization, and replacing $\sigma(t)$ and
$\sigma'(t)$ with symmetrized and antisymmetrized coordinates
$\xi(t)=(\sigma(t)-\sigma'(t))/2$ and
$\eta(t)=(\sigma(t)+\sigma'(t))/2$,
an exact path-integral expression for the transition probability from
position $\sigma_i$ to $\sigma_f$ can be written as
\cite{leggett,weiss}
\begin{eqnarray}
\label{pss}
P_{\sigma_i\sigma_f}(t) &=&
\delta_{\sigma_i, \sigma_f} +\sigma_i\sigma_f
\sum_{n=1}^{\infty}\left(-\frac{\Delta^2}{4} \right)^n \nonumber\\
&&\hspace{-2em}\times\int\limits_{0}^{t} {\cal D}_n\{t_j\}
\sum_{\{\xi_j,\eta_j=\pm 1\}}
\prod_{l=1}^n B_l \prod_{k=1}^{l-1} F_{lk}\;,
\end{eqnarray}
where we have used the shorthand notation
\begin{equation}
\int\limits_{0}^{t} {\cal D}_n\{t_j\} \equiv \int\limits_0^t dt_{2n}
\int\limits_0^{t_{2n}} dt_{2n-1} \ldots \int\limits_0^{t_2} dt_{1}
\end{equation}
for the time-ordered integrals. 
The $\{t_j\}$ here are the times at which transitions are made on the 
$\xi(t)$ and $\eta(t)$ paths and denoting the directions of these 
transitions by $\xi_j$ and $\eta_j$ (called ``charges''), 
the interaction terms $B_l$ and $F_{lk}$ are related to the
real and imaginary part of the twice-integrated bath correlation
function in the following way
{\samepage
\begin{eqnarray}
S(t) + iR(t) &=& \frac{a^2}{\pi\hbar}\int\limits_0^{\infty}d\omega
\frac{J(\omega)}{\omega^2}\Bigg[(1-\cos\omega t)\coth\left(
\frac{\hbar\beta\omega}{2} \right) \nonumber \\
&&
\hspace{10em} +\, i \sin\omega t \Bigg]
\end{eqnarray}}
through
\begin{eqnarray}
B_l &=& \exp( -\xi_{2l-1}\xi_{2l}S_{2l,2l-1} 
       +i\xi_{2l-1}\epsilon(t_{2l}-t_{2l-1}) \nonumber \\
    && +i\xi_{2l} \eta_{2l-1} R_{2l,2l-1})\;, 
\end{eqnarray}
and
\begin{eqnarray}
F_{lk} &=& \exp\left(-\xi_{2l}\xi_{2k-1}S_{2l,2k-1} \right. 
          -\xi_{2l-1}\xi_{2k}S_{2l-1,2k} \nonumber \\
       && -\xi_{2l}\xi_{2k}S_{2l,2k} 
          \left. -\xi_{2l-1}\xi_{2k-1}S_{2l-1,2k-1}  \right) \nonumber \\
       & & \exp\left( i\xi_{2l}\eta_{2k-1}R_{2l,2k-1} \right. 
          +i\xi_{2l-1}\eta_{2k-1}R_{2l-1,2k-1}) \nonumber \\
       && +i\xi_{2l}\eta_{2k}R_{2l,2k} 
          \left. +i\xi_{2l-1}\eta_{2k}R_{2l-1,2k} \right)\;,
\end{eqnarray}
where $S_{mn} \equiv S(t_m - t_n)$ and $R_{mn} \equiv R(t_m - t_n)$.
Because of the two-state nature of the electron coordinate, 
the charges are constrained such that 
$\xi_{2l-1} = -\xi_{2l}$ and $\eta_{2l} = -\eta_{2l+1}$.
These constraints imply that there are two
kinds of time intervals on the double path -- 
those intervals with $\sigma=\sigma'$ (dubbed `sojourns' by Leggett et
al. \cite{leggett}) and those with $\sigma=-\sigma'$ (called `blips').
The double path alternates between sojourns and blips and the expression 
(\ref{pss}) for $P_{\sigma_i\sigma_f}(t)$
involves pair interactions between the charges.  
In (\ref{pss}), we have deliberately broken up the charge-charge interactions 
into ``intrablip'' terms ($B$) and ``interblip'' terms ($F$).

All of the transition probabilities $P_{\sigma_i\sigma_f}(t)$
are related to each other by the conservation of probability.  By
symmetry they can be obtained from a single quantity
\begin{equation}
P(t) \equiv \langle\sigma_z(t)\rangle \equiv P_{+,+}(t) - P_{+,-}(t)\;.
\label{pt}
\end{equation}
After some (possibly complicated) initial transient, 
$P(t)$ should follow a simple exponential decay 
with a inverse time constant equal to the {\em net} reaction rate $\Gamma$.

\section{Dynamical rate theory for electron transfer}
\label{dyntransfer}
In this section, we will discuss the meaning of (\ref{pss}) using 
diagrams.  We begin with the nonadiabatic limit and proceed to the
more general case at the end of the section.

\subsection{Dynamical rate expression in the nonadiabatic limit}
\label{nonadrate}
In the limit of {\em nonadiabatic\/} electron transfer, the
infinite series in (\ref{pss}) can readily be resummed using the
noninteracting blip approximation (NIBA) \cite{leggett,weiss}.  
Because we shall use this later as the basis of an exact expression 
for the reaction rate, we first discuss in detail the diagrammatic
treatment of the nonadiabatic limit.

In the NIBA expression $P'(t)$, the factors $F_{lk}$, which denote
interactions among charges belonging to different blips,
are approximated by unity. 
This turns the multiple integrations in (\ref{pss}) into
convolution products, which can 
\widetext
\vbox{
\begin{figure}
\epsfigrotl{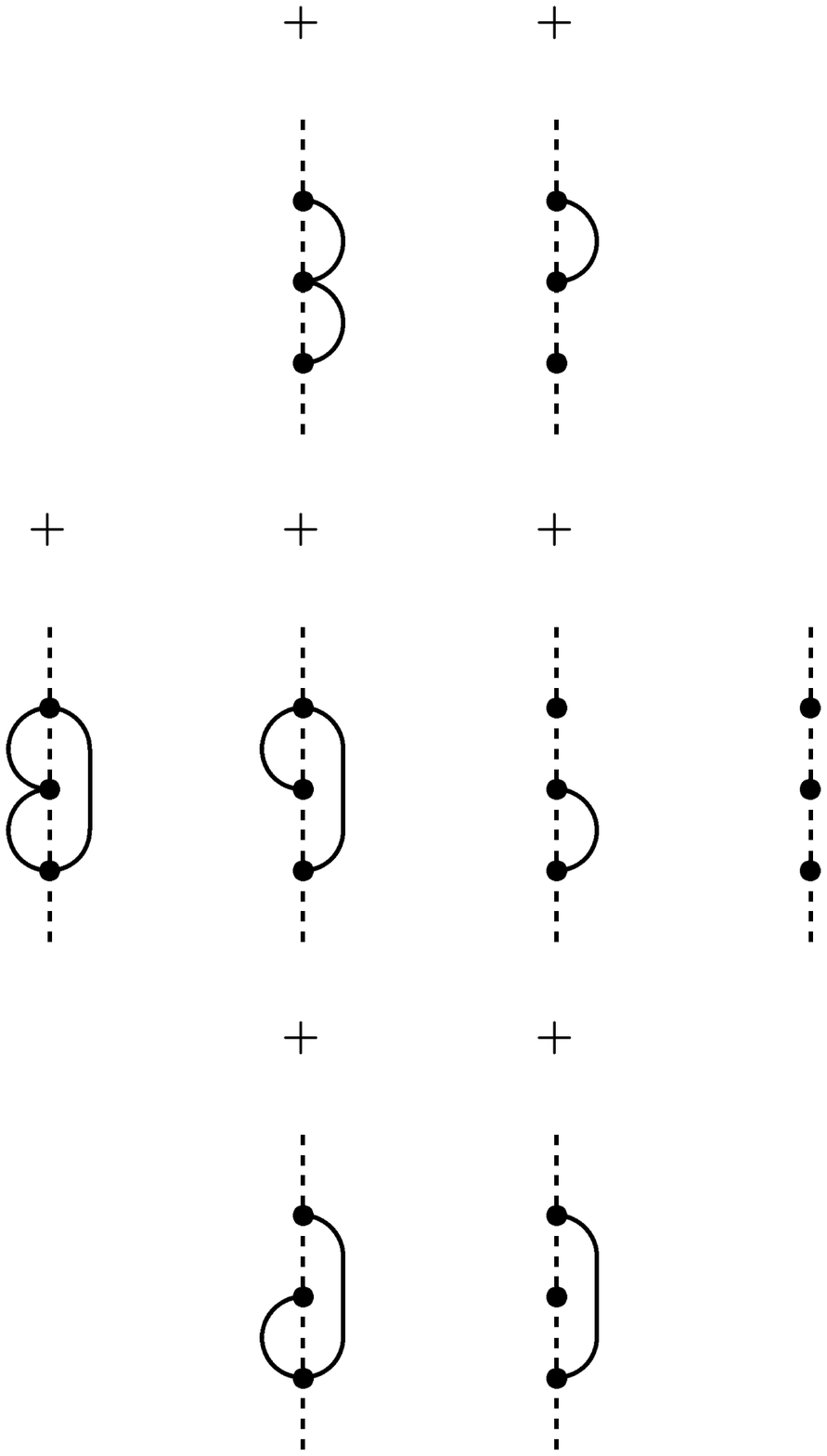}{3.00in}{0.0in}{100 0 496 842}
\caption[]
{\label{diagrams}
Diagrammatic representation of blip-blip interaction terms
$G_{jk}$ for $n=3$. Nodes represent blips, straight lines sojourns,
and arches interaction terms $G_{jk}$. The last five diagrams are irreducible,
while the first three are not.
}
\end{figure}
\narrowtext
}
be Laplace transformed.
The infinite sum in (\ref{pss}) then becomes a geometric
series of the Laplace variable $\lambda$
\begin{equation}
p'(\lambda) = {1\over\lambda}\left(1+{\Sigma_s^{(1)}
 + \Sigma_a^{(1)}
\over \lambda - \Sigma_s^{(1)}}\right)\;,
\label{plniba}
\end{equation}
where
\begin{eqnarray}
\Sigma_s^{(1)}(\lambda) &=& -{\Delta^2\over 4}\int\limits_0^\infty d \tau
\sum_{\xi,\eta=\pm 1}
\exp\left[-\lambda\tau-S(\tau) \right. \nonumber \\
&& \hspace{6em} \left. +i\xi(\epsilon\tau+\eta
R(\tau))\right]\;,\\
\Sigma_a^{(1)}(\lambda) &=& -{\Delta^2\over 4}\int\limits_0^\infty d \tau
\sum_{\xi,\eta=\pm 1}
\eta
\exp\left[-\lambda\tau-S(\tau) \right. \nonumber \\
&& \hspace{6em} \left. +i\xi(\epsilon\tau+\eta
R(\tau))\right]\;.
\end{eqnarray}
In the long-time limit, eq. (\ref{plniba}) gives an exponential decay
when transformed back. The net reaction rate $\Gamma$ is
obtained as the smallest real solution of the equation
\begin{equation}
\label{nibapol}
\Gamma + \Sigma_s^{(1)}(-\Gamma) = 0 \;.
\end{equation}
This result is closely related to the golden-rule rate
\begin{equation}
\label{gr}
\Gamma_{\scsc gr} \equiv - \Sigma_a^{(1)}(\lambda=0)\;,
\end{equation}
which is just a first approximation to iterative solution of
(\ref{nibapol}).

Diagrammatically, we can represent this approximation by the 
following series of graphs

\break{3.25in}
\begin{eqnarray}
\label{nibadiag}
\begin{picture}(200,120)(0,250)

\put(10,353){\line(1,0){32}}
\put(50,350){=}
\multiput(70,353)(4,0){8}{\line(1,0){2}}
\put(110,350){+}
\multiput(130,353)(4,0){6}{\line(1,0){2}}
\put(154,353){\circle*{4}}
\multiput(154,353)(4,0){6}{\line(1,0){2}}

\put(50,300){+}
\multiput(70,303)(4,0){6}{\line(1,0){2}}
\put(94,303){\circle*{4}}
\multiput(94,303)(4,0){6}{\line(1,0){2}}
\put(118,303){\circle*{4}}
\multiput(118,303)(4,0){6}{\line(1,0){2}}

\put(50,250){+}
\multiput(70,253)(4,0){6}{\line(1,0){2}}
\put(94,253){\circle*{4}}
\multiput(94,253)(4,0){6}{\line(1,0){2}}
\put(118,253){\circle*{4}}
\multiput(118,253)(4,0){6}{\line(1,0){2}}
\put(142,253){\circle*{4}}
\multiput(142,253)(4,0){6}{\line(1,0){2}}
\put(190,250){+}
\put(210,250){$\cdots$}

\end{picture}
\nonumber\\
\end{eqnarray}
The solid line in (\ref{nibadiag}) denotes the NIBA transition probability
$P'_{\sigma\sigma'}$, which is equal to a sum over diagrams
representing all possible alternating sequences of sojourns and blips.
Dashed lines represent sojourns,
each of which contributes a factor of unity.  
Dots represent blips, each of which contributes a factor of
$\pm(\Delta/2)^2 B_l$, corresponding to two electronic transitions and
the interactions of the associated charges.
For each internal sojourn line the summation over its $\eta$ label is
implied. Equally, each node also represents a summation over the
blip's $\xi$ label and integration over the two transition times.
Switching to the Laplace transform eliminates one of these
integrations, and changes the other into an integration over the blip
time\cite{funusual} $\tau_l = t_{2l} - t_{2l-1}$. These integrals
factorize, allowing the resummation (\ref{plniba}).

\subsection{Exact dynamical rate expression}
\label{exactrate}
We now turn to an exact diagrammatic representation for $P(t)$.
Using NIBA as the zero-order solution, we develop a perturbation series
for $P(t)$.
In place of the NIBA $F_{lk}=1$, employing the ansatz
\begin{equation}
F_{lk} = 1 + G_{lk} \label{fe1pg}
\end{equation}
expands the product $\prod F_{lk}$ in (\ref{pss}) into a sum,
which is conveniently represented by another series of diagrams.
To generate all terms in this series, we take 
(\ref{nibadiag}) and add to it all diagrams 
in which each pair $(l,k)$ of nodes $l \neq k$ may now be connected by 
(at most) one curved line that represents the interaction $G_{jk}$.
As an example, all third order diagrams generated this way are depicted
in Fig.~\ref{diagrams}.

Some of these diagrams are {\rm reducible\/} in the sense that they
can be separated into disjoint parts by removing a sojourn line.
(An isolated single node is taken as the smallest connected diagram). 
Every reducible diagram can thus be uniquely decomposed into a
sequence of irreducible diagrams and sojourn lines. The {\em
linked-cluster theorem}\/ similarly states that the sum of
all reducible diagrams is graphically represented by
\begin{eqnarray}
\label{resummed}
\begin{picture}(200,70)(0,300)

\put(10,353){\line(1,0){32}}
\put(10,355){\line(1,0){32}}
\put(50,350){=}
\multiput(70,353)(4,0){8}{\line(1,0){2}}
\put(110,350){+}
\multiput(130,353)(4,0){6}{\line(1,0){2}}
\put(160,353){\circle{12}}
\multiput(166,353)(4,0){6}{\line(1,0){2}}

\put(50,300){+}
\multiput(70,303)(4,0){6}{\line(1,0){2}}
\put(100,303){\circle{12}}
\multiput(106,303)(4,0){6}{\line(1,0){2}}
\put(136,303){\circle{12}}
\multiput(142,303)(4,0){6}{\line(1,0){2}}
\put(186,300){+}
\put(206,300){$\cdots$}

\end{picture}
\nonumber \\
\end{eqnarray}
where the linked-cluster sum (or self-energy, in the language of
quantum field theory) represented by the open circle is
\begin{eqnarray}
\label{selfenergy1}
\begin{picture}(200,270)(0,100)

\put(20,353){\circle{12}}
\put(50,350){=}
\put(90,353){\circle*{4}}

\put(50,300){+}
\multiput(70,303)(4,0){6}{\line(1,0){2}}
\put(70,303){\circle*{4}}
\put(94,303){\circle*{4}}
\put(82,303){\oval(24,24)[t]}
\put(110,300){+}
\multiput(130,303)(4,0){6}{\line(1,0){2}}
\multiput(154,303)(4,0){6}{\line(1,0){2}}
\put(130,303){\circle*{4}}
\put(154,303){\circle*{4}}
\put(178,303){\circle*{4}}
\put(154,303){\oval(48,24)[t]}

\put(50,250){+}
\multiput(70,253)(4,0){6}{\line(1,0){2}}
\multiput(94,253)(4,0){6}{\line(1,0){2}}
\multiput(118,253)(4,0){6}{\line(1,0){2}}
\put(70,253){\circle*{4}}
\put(94,253){\circle*{4}}
\put(118,253){\circle*{4}}
\put(142,253){\circle*{4}}
\put(106,253){\oval(72,24)[t]}
\put(160,250){+}
\put(180,250){$\cdots$}

\put(50,200){+}
\multiput(70,203)(4,0){6}{\line(1,0){2}}
\multiput(94,203)(4,0){6}{\line(1,0){2}}
\put(70,203){\circle*{4}}
\put(94,203){\circle*{4}}
\put(118,203){\circle*{4}}
\put(82,203){\oval(24,24)[t]}
\put(106,203){\oval(24,24)[t]}
\put(128,200){+}
\multiput(150,203)(4,0){6}{\line(1,0){2}}
\multiput(174,203)(4,0){6}{\line(1,0){2}}
\put(150,203){\circle*{4}}
\put(174,203){\circle*{4}}
\put(198,203){\circle*{4}}
\put(162,203){\oval(24,24)[t]}
\put(186,203){\oval(24,24)[t]}
\put(174,203){\oval(48,24)[b]}

\put(50,150){+}
\multiput(70,153)(4,0){6}{\line(1,0){2}}
\multiput(94,153)(4,0){6}{\line(1,0){2}}
\put(70,153){\circle*{4}}
\put(94,153){\circle*{4}}
\put(118,153){\circle*{4}}
\put(94,153){\oval(48,24)[t]}
\put(82,153){\oval(24,24)[b]}
\put(128,150){+}
\multiput(150,153)(4,0){6}{\line(1,0){2}}
\multiput(174,153)(4,0){6}{\line(1,0){2}}
\put(150,153){\circle*{4}}
\put(174,153){\circle*{4}}
\put(198,153){\circle*{4}}
\put(174,153){\oval(48,24)[t]}
\put(186,153){\oval(24,24)[b]}

\put(50,100){+}
\multiput(70,103)(4,0){6}{\line(1,0){2}}
\multiput(94,103)(4,0){6}{\line(1,0){2}}
\multiput(118,103)(4,0){6}{\line(1,0){2}}
\put(70,103){\circle*{4}}
\put(94,103){\circle*{4}}
\put(118,103){\circle*{4}}
\put(142,103){\circle*{4}}
\put(106,103){\oval(72,24)[t]}
\put(106,103){\oval(24,24)[b]}
\put(160,100){+}
\put(180,100){$\cdots$}

\end{picture}
\nonumber\\
\end{eqnarray}
In (\ref{resummed}), the double line represents the exact $P(t)$
and dashed lines are again sojourns.
The smallest irreducible diagram in (\ref{selfenergy1}) is just a single node
(i.e. the right-hand side of the first line of (\ref{selfenergy1}), 
representing a blip that does not interact with anything else.

In this decomposition each of 
the multiple integral in (\ref{pss}) in the Laplace transform
$p(\lambda)$ becomes a convolution
product of several terms, each corresponding to an irreducible
self-energy diagram. Formally, applying the linked-cluster theorem means
rearranging the summations
in (\ref{pss}) and (\ref{fe1pg}) to make full use of this
property of the integrand. The new outermost summation is then taken over
the number of connected subdiagrams and leads again to a geometric series
for $p(\lambda)$,
\begin{equation}
\label{pl}
p(\lambda) = {1\over\lambda}\left(1+{\Sigma_s + \Sigma_a
\over \lambda - \Sigma_s}\right) \;.
\end{equation}
The self-energies 
\begin{equation}
\Sigma_s(\lambda) = {1\over 2}\sum_{\eta_0=\pm 1} \Sigma_{\eta_0}\;,\;\;
\Sigma_a(\lambda) = {1\over 2}\sum_{\eta_0=\pm 1} \eta_0 \Sigma_{\eta_0}
\end{equation}
are the symmetric and antisymmetric part of the sum over all
irreducible diagrams, and 
\begin{eqnarray}
\Sigma_{\eta_0}(\lambda) &=&
\sum_{n=1}^\infty \left(\!-{\Delta^2\over 4}\right)^n
\int\limits_{0}^{\infty}\! d\tau^n \!\int\limits_{0}^{\infty} \!ds^{n-1}
\!\!\!\sum_{\{\xi_j,\eta_j=\pm 1\}}\!\!
 B_1 e^{-\lambda\tau_1} \nonumber\\
&&\hspace{-2em}\times\prod_{j=1}^{n-1} e^{-\lambda(s_j+\tau_{j+1})}\, B_{j+1}
\sum_{{\cal C}\in {\sf C}_n}\prod_{(j,k)\in {\cal C}} G_{jk}  \;.
\label{selfsum1}
\end{eqnarray}
Here ${\sf C}_n$ denotes the set of all irreducible diagrams with $n$
nodes, and the notation $(j,k)\in {\cal C}$ means that the graph $\cal
C$ contains a line connecting nodes $j$ and $k$. The integration
variables have been changed to the blip times $\tau_j=t_{2j}-t_{2j-1}$
and sojourn times $s_j=t_{2j+1}-t_{2j}$, where $t_1$ now marks the
beginning of an irreducible cluster.

In this expression, an irreducible diagram may contain any number of
internal nodes that are not part of any interaction line. Such
subdiagrams enclosed by two `interacting' nodes may be summed over in
a fashion similar to NIBA and treated as an `extended sojourn'.
Equation (\ref{selfenergy1}) then assumes the equivalent form
\cite{fWW}
\begin{eqnarray}
\label{selfenergy2}
\begin{picture}(200,170)(0,200)

\put(20,353){\circle{12}}
\put(50,350){=}
\put(70,353){\circle*{4}}
\put(90,350){+}
\put(110,353){\line(1,0){24}}
\put(110,353){\circle*{4}}
\put(134,353){\circle*{4}}
\put(122,353){\oval(24,24)[t]}

\put(50,300){+}
\put(70,303){\line(1,0){24}}
\put(94,303){\line(1,0){24}}
\put(70,303){\circle*{4}}
\put(94,303){\circle*{4}}
\put(118,303){\circle*{4}}
\put(82,303){\oval(24,24)[t]}
\put(106,303){\oval(24,24)[t]}
\put(130,300){+}
\put(150,300){$\cdots$}

\put(50,250){+}
\put(70,253){\line(1,0){24}}
\put(94,253){\line(1,0){24}}
\put(70,253){\circle*{4}}
\put(94,253){\circle*{4}}
\put(118,253){\circle*{4}}
\put(94,253){\oval(48,24)[t]}
\put(82,253){\oval(24,24)[b]}
\put(128,250){+}
\put(150,253){\line(1,0){24}}
\put(174,253){\line(1,0){24}}
\put(150,253){\circle*{4}}
\put(174,253){\circle*{4}}
\put(198,253){\circle*{4}}
\put(174,253){\oval(48,24)[t]}
\put(186,253){\oval(24,24)[b]}

\put(50,200){+}
\put(70,203){\line(1,0){24}}
\put(94,203){\line(1,0){24}}
\put(70,203){\circle*{4}}
\put(94,203){\circle*{4}}
\put(118,203){\circle*{4}}
\put(82,203){\oval(24,24)[t]}
\put(106,203){\oval(24,24)[t]}
\put(94,203){\oval(48,24)[b]}
\put(128,200){+}
\put(150,200){$\cdots$}

\end{picture}
\nonumber \\
\end{eqnarray}
To obtain this series, we eliminate from 
(\ref{selfenergy1}) all contiguous sequences of unconnected 
internal blips and
replace each sequence by a single NIBA (solid) line.
For example, 
the first line in (\ref{selfenergy2}) in the result of resumming
the first three lines in (\ref{selfenergy1}).
Mathematically, this translates to 
\begin{eqnarray}
\Sigma_{\eta_0}(\lambda) &=&
\sum_{{\cal C} \in {\sf G}}
\left(-{\Delta^2\over 4}\right)^{n({\cal C})}
\int\limits_{0}^{\infty} d\tau^n \int\limits_{0}^{\infty} ds^{n-1} \nonumber\\
&& \sum_{\{\xi_j,\eta_j=\pm 1\}} B_1 e^{-\lambda\tau_1}
\prod_{j=1}^{n({\cal C})-1}
\sum_{\eta'_j=\pm 1} \eta_j\eta'_j P'_{\eta_j\eta'_j}(s_j)
\nonumber \\
&& \times e^{-\lambda(s_j+\tau_{j+1})} B'_{j+1}
\prod_{(j,k)\in {\cal C}} G_{jk}  \label{selfsum2}\;,
\end{eqnarray}
where $n(\cal C)$ is the number of nodes in a graph $\cal C$, and
$P'_{\sigma_i,\sigma_f}(t)$ is the NIBA transition probability.
The terms $B'_j$ differ from $B_j$ by the substitution $\eta_{j} \to
\eta'_j$.
It is important to note that the set $\sf G$ contains only
irreducible diagrams with a further restriction: Except in the
trivial single-blip diagram, each node must be part of an interaction
line
\cite{fPreno}.

The reaction rate $\Gamma$ can now be obtained from either
(\ref{selfsum1}) or (\ref{selfsum2}) as the solution of
\begin{equation}
\Gamma + \Sigma_s(-\Gamma) = 0 \;.                     \label{gammeq}
\end{equation}
For relaxation rates slow compared to the dynamical timescales of
the system, (\ref{gammeq}) reduces to
\begin{equation}
\Gamma = \Sigma_s(\lambda=0).
\end{equation}

With (\ref{selfsum1}) and (\ref{selfsum2}) we have thus found a rigorous
result for the dynamics of the spin-boson system. Together
with reasonable approximations for the self-energy, we can produce 
systematic improvements on the NIBA.

\subsection{Leading adiabatic corrections: breakdown of
steepest descent}
\label{adcorr}
Armed with these formulae, we can study the adiabatic corrections,
i.e. deviations from the $O(\Delta^2)$ NIBA. A startling feature of
the adiabatic electron transfer problem is the fact that the Gaussian
approximation for the integrand in (\ref{selfsum1}) obtained by a
high-temperature expansion of the correlation functions $S$ and $R$
breaks down for terms {\em of any order higher than} $\Delta^2$. A
simple examination of the fourth-order term will reveal the problem.
Assuming for the sake of argument that temperatures in the classical
regime $kT\gg\hbar\omega_c$ justify an expansion of the functions
$S(t)$ and $R(t)$, the fourth-order term of $\Sigma_{\eta_0}$ takes
the form
\begin{eqnarray}
\Sigma_{\eta_0}^{(2)} &=& {\Delta^4\over 8}
\sum_{\xi_1,\eta_1,\xi_2=\pm 1}\,
\int\limits_0^\infty \!d\tau_1
\int\limits_0^\infty \!d s_1
\int\limits_0^\infty \!d\tau_2 \,
e^{-\lambda\tau_1 -\lambda s_1 -\lambda\tau_2}
\nonumber\\
&&\hspace{-2em}\times \exp(-\Lambda kT\tau_1^2/\hbar +
i(\epsilon+\eta_0\Lambda)\xi_1\tau_1 + {\cal R}_1)
\nonumber\\
&&\hspace{-2em}\times \exp(-\Lambda kT\tau_2^2/\hbar +
i(\epsilon+\eta_1\Lambda)\xi_2\tau_2 + {\cal R}_2)
\nonumber\\
&&\hspace{-2em}\times\big[\exp(-2\xi_1\xi_2\Lambda kT\tau_1\tau_2/\hbar
   +i(\eta_0-\eta_1)\xi_2\Lambda\tau_2 + {\cal R}_I)
\nonumber\\
&& - 1 \big]\;.
\end{eqnarray}
The terms ${\cal R}_1$, ${\cal R}_2$, and ${\cal R}_I$ denote
contributions of higher than quadratic order in the exponent, and
$\hbar\Lambda=2\alpha\,\hbar\omega_c$ is the reorganization energy of the
environment.
With $\tilde \tau_j =
\xi_j\tau_j$ and $\tilde s_j = \eta_j s_j$, the integrations in
\begin{eqnarray}
\Sigma_{\eta_0}^{(2)} &=&
\Delta^4 \int\limits_{-\infty}^\infty d \tilde s_1 e^{-\lambda|\tilde s_1|}
\Bigg\{ \int\limits_{-\infty}^\infty d \tilde \tau_1
e^{-\lambda|\tilde \tau_1|}
        \int\limits_{-\infty}^\infty d \tilde \tau_2
e^{-\lambda|\tilde \tau_2|}
\nonumber\\
&&\hspace{-2em}
\times \exp\left[\Lambda kT(\tilde\tau_1+\tilde\tau_2)^2/\hbar
     +i(\eta_0\Lambda+\epsilon)(\tilde\tau_1+\tilde\tau_2)+\bar{\cal
     R}\right]
\nonumber\\
&& \hspace{12em} - \Sigma_{\eta_0}^{(1)} \Sigma_s^{(1)}
\Bigg\}                                           \label{sig2}
\end{eqnarray}
become formally Gaussian for $\lambda,\bar{\cal R}\to 0$. Due to the
degeneracy of the quadratic form in the exponent, however, one is {\em
forced\/} to retain the higher-order corrections $\bar{\cal R}$ in
order to prevent the integral from diverging. The same picture,
with the quadratic part of the exponent depending only on one single
variable $\tau^* = \sum_j
\xi_j \tau_j$ and $2n-2$ `false' zero modes, emerges for terms of
arbitrary order $\Delta^{2n}$.
Thus, it is not surprising that a numerical evaluation of the complete
expression (\ref{sig2}) leads to results quite different from those
obtained when approximating the bath correlation function by its
short-time behavior.

Following Garg et al. \cite{garg}, a convenient
quantitative description of the transition from nonadiabatic to
adiabatic behavior can be given in terms of an {\em adiabaticity
factor}\/ $g(\Delta,\epsilon,\omega_c,\alpha,T)$ defined by
\begin{equation}
g = - {\Gamma - \Gamma_{\scsc gr} \over \Gamma} \equiv - \left.
{\Sigma_s - \Sigma_s^{(1)} \over \Sigma_s} \right|_{\lambda=0}\;,
\label{adfac}
\end{equation}
which specifies the relative weight of the non-trivial diagrams in
(\ref{selfenergy1}), i.e., the multiblip contributions to
(\ref{selfsum1}). When adiabaticity effects {\em reduce} the rate, the
sum associated with these higher-order diagrams can exceed the
rate itself in absolute value, making $g$ a {\em positive}
number. A {\em negative} $g$, on the other hand, would indicate higher
order terms {\em promoting} electron transfer.

Restricting (\ref{selfsum1}) to terms of order $\Delta^4$ for
small adiabatic corrections $g\ll 1$, (\ref{adfac}) turns into
\begin{equation}
g = 1 - {\Gamma \over \Gamma_{\scsc gr}}\;,
\end{equation}
and $g$ scales as
\begin{equation}
g = {\Delta^2\over \omega_c^2}
\tilde{g}(\alpha,kT/\hbar\omega_c,\epsilon/\omega_c)\;.
\end{equation}
We have evaluated $\tilde{g}(\alpha,kT/\hbar\omega_c)$ numerically for a
symmetric system, with results shown in Fig.~\ref{advsa}. For
high temperatures $kT \gg \hbar\Lambda$ the adiabaticity factor is only
weakly dependent on $\alpha$, but still varies significantly with
temperature. Here we are in disagreement with Ref. \cite{garg}, where a
temperature-independent adiabaticity factor $g\propto \alpha^{-1}$ was found
\cite{fDrude}.

At lower temperatures, but still with $kT>\hbar\omega_c$ we find that
the function $\tilde{g}(\alpha,kT/\hbar\omega_c)$ obeys a further approximate
scaling relation
\begin{equation}
\tilde{g}(\alpha,kT/\hbar\omega_c) \propto \left(\hbar\omega_c\over
kT\right)^\kappa w\left(\hbar\Lambda\over kT\right)\;, \label{adscale}
\end{equation}
where $\kappa\approx 0.8$. The function $w(x)$ approximately follows
an Arrhenius law with a very small activation energy $\approx
0.065\,\hbar\Lambda$. Fig.~\ref{adsc} shows numerical data for the
universal function $w(x)$ with $\alpha$ varying from 4 to 100, and
with temperatures from $kT/\hbar\omega_c=4$ to 80.

Our result (\ref{adscale}) also shows a significant discrepancy
compared to thermodynamic rate expressions as, e.g., in the approach
by Song and Stuchebrukhov \cite{stuchebrukhov} using Langer's
method \cite{Langer}, where the only relevant parameter 
for an environment
of `classical' modes with $\hbar\omega_c\ll kT$ are the reorganization
energy $\hbar\Lambda$ and temperature $T$. In their approach, the
characteristic frequency $\omega_c$ ceases to be a relevant parameter.
The scenario most frequently invoked to justify Langer's
method, however, supposes that the action can be factorized near a
saddlepoint, with a unique unstable coordinate being the reaction
coordinate, and other coordinates undergoing only fluctuations. To our
knowledge, this property of the action has not been demonstrated for
adiabatic ET or for the crossover region, nor has the unstable coordinate
been identified.

\subsection{Theory for the crossover from nonadiabatic to adiabatic 
electron transfer}
\label{crossover}
Although approximating (\ref{selfsum1}) by a finite number of diagrams
has provided us with valuable information about the leading
corrections to nonadiabatic theory, a better approximation is needed
for the crossover to adiabatic electron transfer. In the delocalized
transition state, the electron may oscillate back and
forth many times before settling on one of the two sites. Moreover, in
the case of strong dissipation correlated recrossings of the reaction
barrier must be considered for an accurate rate expression.

An exact resummation of (\ref{selfsum1}) or (\ref{selfsum2}) seems out
of reach without further insight into the formal structure of the
problem. We shall present an approximate resummation of
(\ref{selfsum2}) instead, which captures the most important
contributions in several  parameter regions. Our
approximation is guided by the fact that the interaction terms
$G_{jk}$ fall off rapidly when the separation of blips is larger than
$\omega_c^{-1}$, which is typically a much shorter timescale than the
inverse of the reaction rate. For any diagram 
\vbox{
\begin{figure}
\epsfig{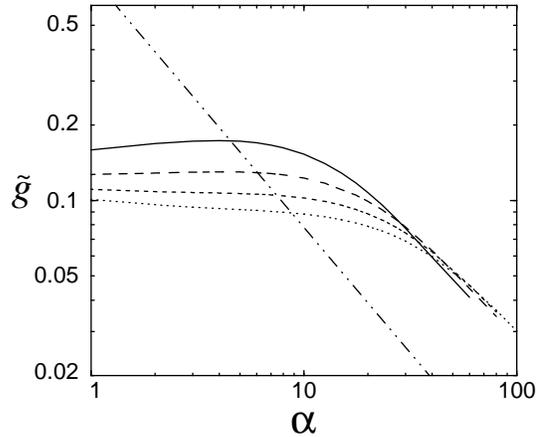}{2.50in}{-0.30in}{0 200 600 620}
\caption[]
{\label{advsa}
Adiabaticity factor $\tilde{g}$ obtained from the leading
corrections to the Golden Rule. Temperature values are $kT/\hbar\omega_c=4$
(solid line), $6$ (long dash) , $8$ (short dash), and $10$ (dotted
line). The adiabaticity factor of Garg et al. \cite{garg}, which is
temperature-independent, is given for comparison (dash-dotted line).
}
\end{figure}
\begin{figure}
\epsfig{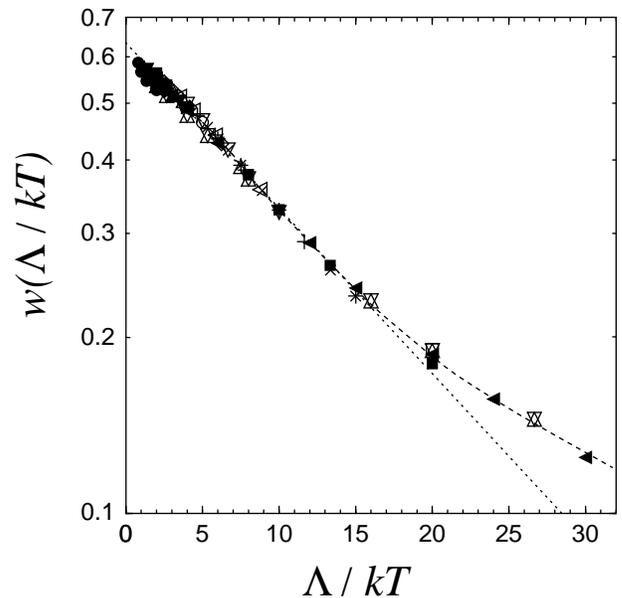}{3.50in}{-0.35in}{0 200 600 750}
\caption[]
{\label{adsc}
Numerical data for the universal function
$w(\hbar\Lambda/kT)$. Temperature varies from $kT/\hbar\omega_c=4$ to
$80$, with $\alpha$ ranging from $4$ to $100$. Dotted line: fit by
$a\exp(-b\,\hbar\Lambda/kT)$ with $a=0.63$ and $b=0.065$. The dashed
line interpolates between data points.
}
\end{figure}
}
with {\em nested}\/
interaction lines, this cutoff applies to the {\em sum}\/ of all blip
and sojourn times spanned by the outermost interaction line. This
constraint decreases the volume of the integration domain by a factor $n!$
compared to case of independent cutoffs for each of $n$ integration
variables. Diagrams with nested lines will therefore not be taken into
account, and diagrams with crossed lines will similarly be neglected.
Therefore, in the self-energy we only retain a set of `bridge diagrams', 
which correspond to only the first two lines of (\ref{selfenergy2}).

It is now useful to keep the integration over the last blip time
$\tau$ and the summation over its indices $\xi$ and $\eta$ separate
from the preceding blips. The self-energy terms can then be written as
a power series of the form
\begin{equation}
\label{selfseries}
\Sigma_{\eta_0} = 2 \sum_{n=1}^\infty
\int\limits_0^\infty d\tau\sum_{\xi=\pm 1}
\Delta^{2n}\tilde\B_n (\tau,\xi,\eta_0)\;.
\end{equation}
For the bridge diagrams, the terms $\tilde\B_n$ obey a linear
recursion relation
\begin{eqnarray}
\label{recursion}
\tilde\B_{n+1}(\tau,\xi,\eta) &=& \Delta^2
\int\limits_0^\infty d\tau' \!\sum_{\xi',\eta'=\pm 1}\!
K(\tau,\xi,\eta;\tau',\xi',\eta')
\nonumber \\
&& \hspace{6em} \times \tilde\B_n (\tau',\xi',\eta')\;,
\end{eqnarray}
where
\begin{eqnarray}
\tilde\B_1 (\tau,\xi,\eta) &=& -{1\over 4}e^{-\lambda\tau}\!
\exp\!\left[-S(\tau)\!+\!i\xi(\epsilon\tau\!+\!\eta R(\tau))\right]\,.\!\!
\end{eqnarray}
The kernel of the integral transform in (\ref{recursion}) is given by
\begin{eqnarray}
\label{kernel}
K(\tau,\xi,\eta;\tau',\xi',\eta') &=&
\tilde\B_1 (\tau,\xi,\eta)
\int\limits_0^{\infty} ds e^{-\lambda s} \nonumber \\
&& \hspace{-7em} \times \sum_{\eta''=\pm 1} \eta\eta''
P'_{\eta\eta''}(s) \,
G(\tau,\xi;\tau',\xi',\eta',\eta'';s)
\end{eqnarray}
where
\begin{eqnarray}
G(\tau,\xi;\tau'\!,\xi'\!,\eta'\!,\eta''\!;s) &=&
\exp [i\eta'\xi(R(\tau'\!\!+\!s+\!\tau)-R(\tau'\!\!+s))]\nonumber\\
&\hspace{-3em}\times\hspace{3em}&\hspace{-3em}
\exp [-i\eta''\xi(R(\tau'+s)-R(s))]\nonumber\\
&\hspace{-3em}\times\hspace{3em}&\hspace{-3em}\exp[-\xi\xi'(
S(\tau'+s+\tau)+S(s) \nonumber \\
&& \hspace{-3em}-S(s+\tau)-S(\tau'+s)
)] \,-\, 1\;.
\end{eqnarray}
>From this recursion relation an integral
equation for the entire series
\begin{equation}
Y (\tau,\xi,\eta)
 = \sum_{n=1}^\infty \Delta^{2n}\tilde\B_n (\tau,\xi,\eta)
\end{equation}
is easily derived,
\begin{eqnarray}
\label{inteq}
&& \hspace{-0.3em} Y(\tau,\xi,\eta) - \Delta^2\tilde\B_1(\tau,\xi,\eta)
\nonumber \\
&& = \Delta^2 \!\int\limits_0^\infty \! d\tau' 
\!\!\! \sum_{\xi',\eta'=\pm 1} \!
K(\tau,\xi,\eta;\tau',\xi',\eta') Y(\tau',\xi',\eta')\;.
\end{eqnarray}
Finally, the linked-cluster sum $\Sigma_{\eta_0}$ is obtained by
applying the final integration and summation
\begin{equation}
\label{finint}
\Sigma_{\eta_0} = 2 \int\limits_0^\infty d\tau \sum_{\xi=\pm 1}
Y(\tau,\sigma,\eta_0)\;.
\end{equation}
With (\ref{kernel}), (\ref{inteq}), and (\ref{finint}) we are now left
with a finite number of quadratures and an integral equation, all of
which can be solved with very modest numerical cost.

A certain degree of control over our approximation of the
linked-cluster sum can also be exercised by the inclusion of a large
class of nested diagrams. This can be done by replacing the NIBA
propagating function $P'_{\eta_j\eta'_j}(s_j)$ in (\ref{selfenergy2}) by
the full expression $P_{\eta_j\eta'_j}(s_j)$ and solving the equation
system given by (\ref{pl}) and (\ref{selfsum2}) self-consistently by
iteration. The results we present in this paper do not change when
these additional diagrams are included.
An appealing aspect of this approach compared to other
dynamical methods is the fact that the dynamics at arbitrarily long
times can be studied by evaluating the linked-cluster
sum at small $\lambda$.

First, in the high-temperature (almost barrierless) case, 
the ET rate should depend on the electron coupling $\Delta/2$ 
only through the adiabaticity prefactor.
In this case, 
electron transfer rates obtained by the diagrammatic method
compare very favorably
with recently published quantum Monte Carlo data \cite{QMC}, as shown in 
Fig.~\ref{gvsd}, 
which shows the dependence of the rate on the electronic coupling.
For the parameters we have chosen here, the activation factor is close
to unity, i.e., the observed variation of the rate is mostly due to
dynamical effects. Our results are in agreement with the traditional
\vbox{
\begin{figure}
\epsfig{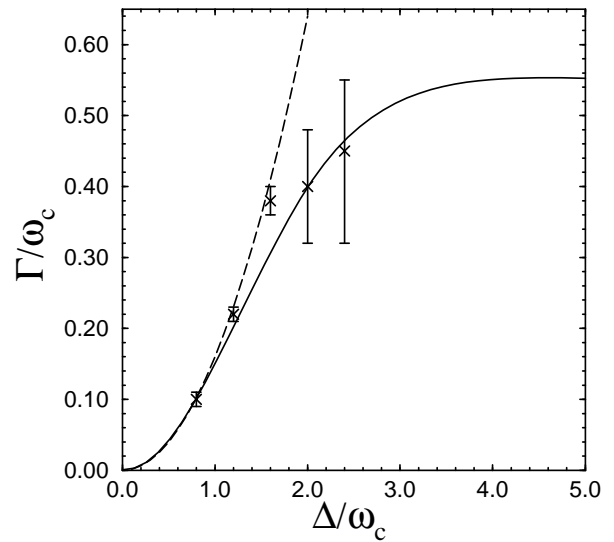}{3.00in}{-0.30in}{0 200 600 720}
\caption[]
{\label{gvsd}
Comparison of the present theory and QMC simulations \cite{QMC} in
the crossover region between nonadiabatic and adiabatic electron
transfer in the high-temperature (almost barrierless) case
for $\Lambda/\omega_c=4$ and
$kT/\hbar\omega_c = 4$. Dashed line: Golden
Rule expectation. Solid line: Summation over bridge diagrams. Symbols: Quantum
Monte Carlo data.
}
\end{figure}
}
\vbox{
\begin{figure}
\epsfig{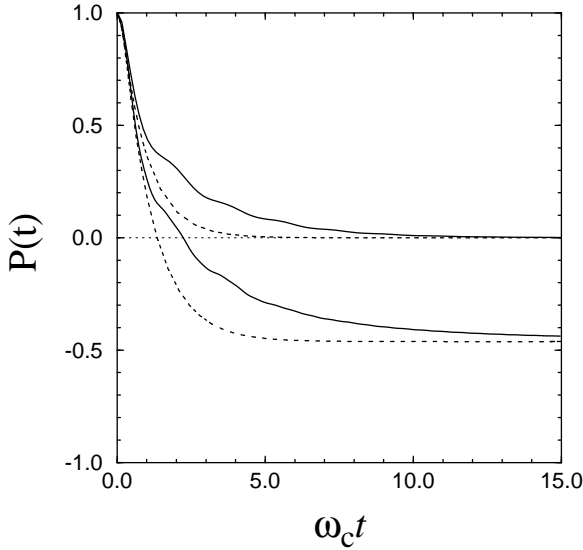}{3.00in}{-0.30in}{0 200 600 720}
\caption[]
{\label{Poft}
Reactant decay for a symmetric self-exchange ET reaction
(zero long-time limit) and for an asymmetric donor-acceptor pair
($\epsilon=4\omega_c$, negative long-time limit) for the same
parameters in Fig.~\ref{gvsd}.  Solid lines represent the resummation
of the bridge diagrams in \ref{selfsum2}); dashed lines give the
noninteracting-blip approximation for comparison.
}
\end{figure}
}
expectation of an initially quadratic dependence on the electronic
coupling that rises and saturates to a plateau
value for very large electronic couplings.

For these parameters, 
as in MC simulations, the time-dependent decay of $P(t)$ shows initial
oscillations before approaching a limit of exponential decay,
Fig.~\ref{Poft}. Note that our method correctly reproduces the
equilibrium value
\begin{equation}
P_\infty = \lim_{t\to \infty} P(t) = \tanh{\hbar \epsilon
\over 2kT}\;,
\end{equation}
i.e., the principle of detailed balance is preserved.

Next, we examine the diagrammatic results in the low-temperature
(activated) region.
In Fig.~\ref{arrh} we present a set of Arrhenius plots for increasing
values of the electronic coupling, but keeping the classical reorganization
energy constant.  At the smallest value
$\Delta/\omega_c=1$, deviations from the NIBA are minute. For somewhat
larger values of $\Delta$, however, one finds significant discrepancies
due to adiabaticity effects.

The detailed features of these deviations conform to a number of known
or intuitively expected characteristics 
of adiabatic electron
transfer. (a) The slope of each Arrhenius plot decreases with
increasing $\Delta$. This of course reflects the lowering of the
activation free energy through the electronic coupling
\cite{gehlen1,gehlen2,stuchebrukhov}.  (b) For high
enough temperatures, the absolute value of the rate is significantly
reduced from the Golden Rule rate, as predicted by dynamical theories
\cite{Zusman,garg}. (c) All results show some
degree of nuclear
tunneling, visible in a deviation from 
\vbox{
\begin{figure}
\epsfig{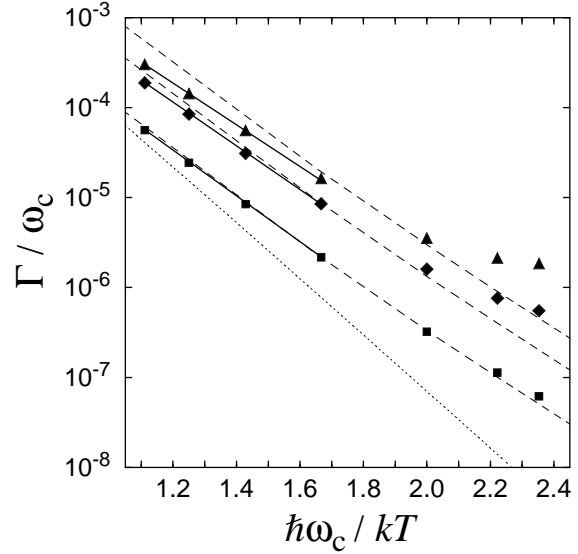}{3.00in}{-0.30in}{0 200 600 720}
\caption[]
{\label{arrh}
Rates obtained in the low-temperature (activated)
case for $\Lambda/\omega_c=30$.
Results from summation over bridge diagrams for
$\Delta/\omega_c=1$ (squares), $\Delta/\omega_c=2$ (diamonds), and
$\Delta/\omega_c=3$ (triangles). The dashed lines indicate the Golden
Rule rate for comparison ($\Delta / \omega_c=1$, $2$, and $3$, bottom
to top). The classical nonadiabatic rate for $\Delta/\omega_c=1$ is
given for comparison (dotted line).  At high enough temperatures,
Arrhenius-like behavior is observed, as indicated by the match between
data and straight solid lines. 
}
\end{figure}
}
Arrhenius behavior at low
temperatures, with a rate that becomes much higher than its expected
classical value.

We can put observations (a) and (b) on a more quantitative basis by
separating the rate into an Arrhenius-like part and a temperature
independent prefactor as
\begin{equation}
\Gamma = \nu \exp(-{F^*/kT})                \label{arrhfit}
\end{equation}
Using (\ref{arrhfit}) as an ansatz to fit our data between
$kT/=0.6\,\hbar\omega_c$ and $kT/=\hbar\omega_c$, we can extract the
$\Delta$-dependence of $\nu$ and $F^*$ separately, as shown in 
Fig.~\ref{nuandF}. The adiabaticity prefactor $\nu$ shows the expected
transition from quadratic to saturation behavior, and the
activation free energy $F^*$ shows approximately the expected linear
decline.

Our diagrammatic theory reproduces the essential
tenets of both the {\em thermodynamic}\/ approach---the modification
the reaction barrier through the electronic coupling---and of {\em
dynamical} theories---the slowing of the reaction due to recrossing
effects---in a {\em unified} description.

\section{Summary and Conclusions}
\label{summary}
In this paper we have presented a novel dynamical approach to the
theory of electron transfer reactions.  The spin-boson model is used
to represent electron transfer in 
\vbox{
\widetext
\begin{figure}
\epsfigrotr{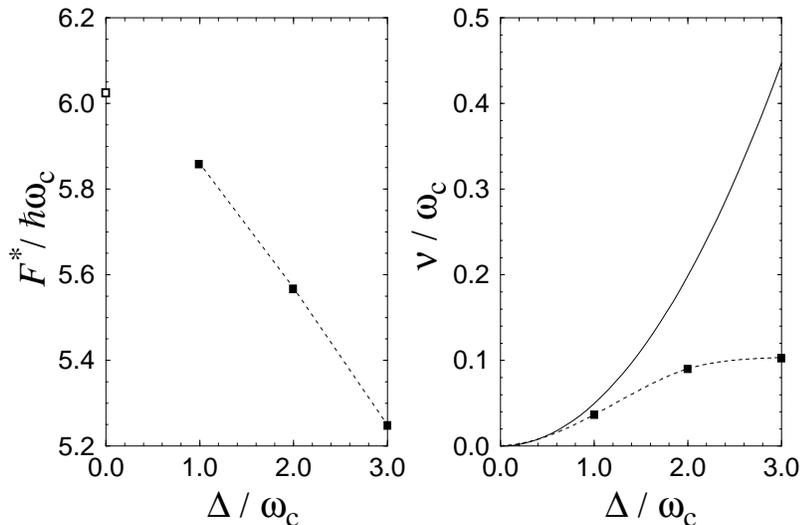}{3.00in}{1.00in}{0 0 546 842}
\caption[]
{\label{nuandF}
(a) Activation free energy
$F^*$ as a function of electronic overlap $\Delta$, using data from
Fig.\ \ref{arrh} between $kT=0.6\,\hbar\omega_c$ and
$kT=\hbar\omega_c$. The open square at $\Delta=0$ denotes the
Golden Rule value. (b) Adiabaticity prefactor $\nu$ from the same data.
}
\end{figure}
\narrowtext
}
condensed-phase from a donor to an
acceptor site. Relying on the established real-time path-integral
representation of the spin-boson dynamics, we have for the first time
rigorously deduced an {\em exact} rate expression in terms of the
spin-boson parameters.  In the nonadiabatic limit, our result reduces
to the simpler NIBA description of spin-boson dynamics. Using the
linked-cluster theorem, we have found a systematic way of summing up
all corrections to the NIBA rate expression. Since we are using a
dynamical treatment directly, our method is valid under rather weak
assumptions and it is applicable in {\em any} parameter region that
would yield a long-time limit of exponential decay of the reactant
population.

We have demonstrated the relevance of our central results and the
continued need for systematic groundwork for the electron transfer
problem by pointing
out significant discrepancies between results based on the
linked-cluster sum (\ref{selfsum1}) and previous approximate real-time
or imaginary-time approaches.
We have also shown the usefulness of our approach by presenting a
resummation of the most important terms in the linked-cluster sum
(\ref{selfsum2}). This resummation imposes no approximation on terms
up to order $\Delta^4$, and it retains a large, dominant class of
diagrams for any order of the electronic or environmental coupling.
Our results compare very favorably with published quantum
Monte Carlo data and the results of a refined version of
the resummation method.

The present approach provides a unified treatment of several
characteristic features of adiabatic electron transfer which were
reported earlier, but separately, using incompatible methods. Our
results show two competing adiabatic effects. First, there is a rate
reduction (compared to the nonadiabatic case), which has previously
\break{3.5in}
been described in terms of correlated recrossings of the 
Landau-Zener
region. This effect is dominant at high temperatures. For low enough
temperatures, one sees a transfer rate that is enhanced over the
nonadiabatic result. This reflects the lowering of the reaction
barrier by the electronic coupling and by nuclear tunneling.
When higher demands on computational resources can be tolerated, our
recursive resummation method may be extended to include larger classes
of diagrams if higher accuracy is desired.

\acknowledgements
We wish to thank Reinhold Egger and Uli Weiss for helpful discussions.
This research is supported by NSF under Grants No. CHE-9216221, 
CHE-9257094, CHE-9528121, and by the Camille and Henry Dreyfus
Foundation and the Alfred P. Sloan Foundation.  Computational resources
have been furnished by the IBM Corporation under the SUR grant.


\begin{references}
\bibitem{funusual} This feature, which may strike some readers as
unusual, has its root in the fact that we are dealing with diagrams
representing {\em probabilities} rather than probability amplitudes.
\bibitem{fWW} The leading terms of (\ref{selfsum2}) for $\alpha\ll 1$
were considered earlier in by Weiss and Wollensak \cite{WW} in the
context of the macroscopic quantum coherence problem.
\bibitem{fPreno} The equation (\ref{resummed}) for $P(t)$ 
together with the self-energy series in (\ref{selfenergy2}) 
can actually be further renormalized to give a self-consistent 
equation for the self-energy which can be solved iteratively.
Since this further renormalization does not lead to any substantial
improvement in the final results, we do not consider it explicitly
here.  Also see Sect.~\ref{crossover}.
\bibitem{fDrude} We performed some of our calculation using a Drude cutoff,
which was used in \cite{garg}, with no essential differences
compared to Fig.~\ref{advsa}.
\end{references}
\end{document}